\documentclass{optica-article}

\journal{oe}

\articletype{Research Article}

\usepackage{lineno}
\usepackage{caption}
\usepackage{subcaption}
\usepackage{physics}

\begin{document}

\title{Demonstration of Hong-Ou-Mandel interference in an LNOI directional coupler}

\author{Silia Babel\authormark{1, *}, Laura Bollmers\authormark{1}, Marcello Massaro\authormark{1}, Kai Hong Luo\authormark{1}, Michael Stefszky\authormark{1}, Federico Pegoraro\authormark{1}, Philip Held\authormark{1}, Harald Herrmann\authormark{1}, Christof Eigner\authormark{1}, Benjamin Brecht\authormark{1}, Laura Padberg \authormark{1}, and Christine Silberhorn\authormark{1}} 

\address{\authormark{1} Paderborn University, Integrated Quantum Optics,
Institute for Photonic Quantum Systems (PhoQS), Warburger Str. 100, 33098 Paderborn, Germany}

\email{\authormark{*}silia.babel@upb.de}
\newcommand{\LP}[1]{\textcolor{violet}{\textbf{LP: }#1}}
\newcommand{\FP}[1]{\textcolor{blue}{\textbf{FP: }#1}}
\newcommand{\SB}[1]{\textcolor{red}{\textbf{ToDo: }#1}}
\newcommand{\CE}[1]{\textcolor{orange}{\textbf{CE: }#1}}
\newcommand{\KL}[1]{\textcolor{cyan}{\textbf{KL: }#1}}
\newcommand{\LB}[1]{\textcolor{teal}{\textbf{LB: }#1}}


\begin{abstract}
Interference between single photons is key for many quantum optics experiments and applications in quantum technologies, such as quantum communication or computation. 
It is advantageous to operate the systems at telecommunication wavelengths and to integrate the setups for these applications in order to improve stability, compactness and scalability. 
A new promising material platform for integrated quantum optics is lithium niobate on insulator (LNOI).
Here, we realise Hong-Ou-Mandel (HOM) interference between telecom photons from an engineered parametric down-conversion source in an LNOI directional coupler. The coupler has been designed and fabricated in house and provides close to perfect balanced beam splitting.
We obtain a raw HOM visibility of ($93.5\pm0.7$)~$\%$, limited mainly by the source performance and in good agreement with off-chip measurements.
This lays the foundation for more sophisticated quantum experiments in LNOI.
\end{abstract}

\section{Introduction}

Linear optical networks are the basis of quantum technologies. 
Many important applications of quantum technologies such as quantum key distribution \cite{duan2001long} and boson sampling \cite{zhong2020quantum} were demonstrated using linear optics. 
At the heart of these linear networks lie optical beam splitters (BS) and phase shifters, since it was shown that with these two components any unitary matrix operation can be implemented \cite{reck1994experimental}.\\
A particular interest has been devoted to the study and realisation of integrated optical circuits, since they offer high stability, the possibility of compact devices and high efficiencies, and thus enable scalability in contrast to their free space or bulk counterparts. 
The history of integrated linear networks is summarised in many review articles see, e.g., \cite{wang2020integrated}.\\ 
A new promising material for the implementation of future quantum circuits is lithium niobate on insulator (LNOI) since it combines a high integration density, as known from silicon photonics, with  its unique functionalities of electro-optical modulation and a second order nonlinearity \cite{zhu2021integrated}. 
It consists of a thin-film of lithium niobate (LN) (300~nm to 1000~nm) bonded on a SiO$_2$ layer. 
The light is guided in the LN thin-film and thus it inherits the diverse property portfolio of LN, e.g., its wide transparency window, its large second order non-linearity and electro-optical coefficients \cite{weis1985lithium}. 
Due to its structure with a SiO$_2$ layer underneath and the low dimension of the LN thin-film a strong confinement for rib waveguides can be realised, giving rise not only to the high integration density, but also to a highly improved effective non-linearity and the possibility of dispersion engineering.\\
LNOI has already been employed for the realisation of a variety of devices, such as electro-optic modulators \cite{wang2018integrated,pan2021demonstration,xue2022breaking}, directional couplers \cite{zhang2020polarization, wang2019chemo}, as well as for second harmonic generation \cite{boes2019improved, park2022high,zhang2022second} and spontaneous parametric down-conversion \cite{xin2022spectrally, henry2022correlated, javid2021ultrabroadband}. All of these functionalities combined with the possibility to achieve very low loss waveguides \cite{zhang2017monolithic} illustrate the high potential of LNOI for future photonic quantum technologies.
A key building block of quantum networks is two-photon Hong-Ou-Mandel (HOM) interference.\\
Here, we demonstrate HOM interference of photons at telecommunication wavelengths from a parametric down-conversion source in an LNOI directional coupler. 
We observe an interference visibility of ($93.5\pm0.7$)~$\%$, limited by the source performance and similar to an off-chip measurement. 
This constitutes an important building block for realising optical quantum networks on LNOI and paves the way towards future applications. \\
This paper is structured as follows. 
In \autoref{fund}, we briefly discuss the basics of HOM interference. 
Section \ref{Simulations} details the theory and the modeling of the LNOI directional coupler and we describe the fabrication process in \autoref{Fabrication}. 
In \autoref{Linear} and \autoref{Quantum}, we present the characterisation of the coupler and the HOM measurements, respectively. Section \ref{Conclusion} concludes the paper. 

\section{Fundamentals} \label{fund}

HOM interference describes the bunching of two indistinguishable photons at a balanced, non-polarising beam splitter (BS) where both photons enter from one input port each, but leave them together in one of the outputs.
Typically, the degree of distinguishability between the two photons is continuously changed from completely distinguishable to fully indistinguishable, by introducing a time delay between the photons. 
Monitoring the number of coincidence events between the output ports of the BS as function of delay then yields the famous HOM dip \cite{hong1987measurement}. 
The visibility of this dip is ideally 100~$\%$ but typically limited by the imbalance of the BS and the indistinguishability of the impinging photons.
Here, we compare the dip visibility, which we obtain with our LNOI device, against the visibility, which we observe with an off-chip reference measurement. This allows us to gauge the impact of the LNOI directional coupler on the photon indistinguishability and hence the applicability of LNOI in integrated optical quantum networks.\\
Perfect HOM interference is only observed with a perfectly balanced BS. Unavoidable fabrication intolerances will, however, lead to deviations from a 50:50 splitting. The impact of these on the HOM visibility can be summarised as \cite{matthews2009manipulation}

\begin{equation}
    V_{max} = \frac{2 \cdot \eta \cdot (1-\eta)}{1- 2 \cdot \eta + 2 \cdot \eta^2},
    \label{VEquation}
\end{equation}
where $\eta$ is the reflectivity of the BS.
To fabricate a 50:50 directional coupler, we first carried out numerical simulations to estimate the coupling behaviour. These are discussed in the next section.

\section{Simulations of Waveguide and Directional Coupler Structures} \label{Simulations}

The aim of our simulations is to find a suitable waveguide and directional coupler geometry that meets our needs such as single modeness, short coupling length, a gap width that is compatible with our fabrication capabilities, and a large bandwidth. For the numerical evaluation, we briefly recap the operation principle of a directional coupler.\\
A directional coupler consists of two waveguides that are brought close to one another for a given length, $L_I$, with a small gap given by $G$.
Along the length $L_I$ the coupling takes place via the evanescent fields of the two waveguides.
Mathematically, this is explained by coupled mode theory which yields that the incoupled power oscillates sinusoidally between the two waveguides \cite{Yariv}. 
The coupling depends on the wavelength, the waveguide geometry, the interaction length and the gap and is summarised by the coupling coefficient. 
A user-chosen splitting ratio can be realised by adapting $L_I$ appropriately.\\ 
We first investigated the single modeness of a single waveguide using the \textit{MODE} solver of \textit{Lumerical} \cite{Lumerical}. 
\begin{figure}
\vspace*{-5mm}
     \centering
     \begin{subfigure}[b]{0.3\textwidth}
         \centering
         \includegraphics[width=\textwidth]{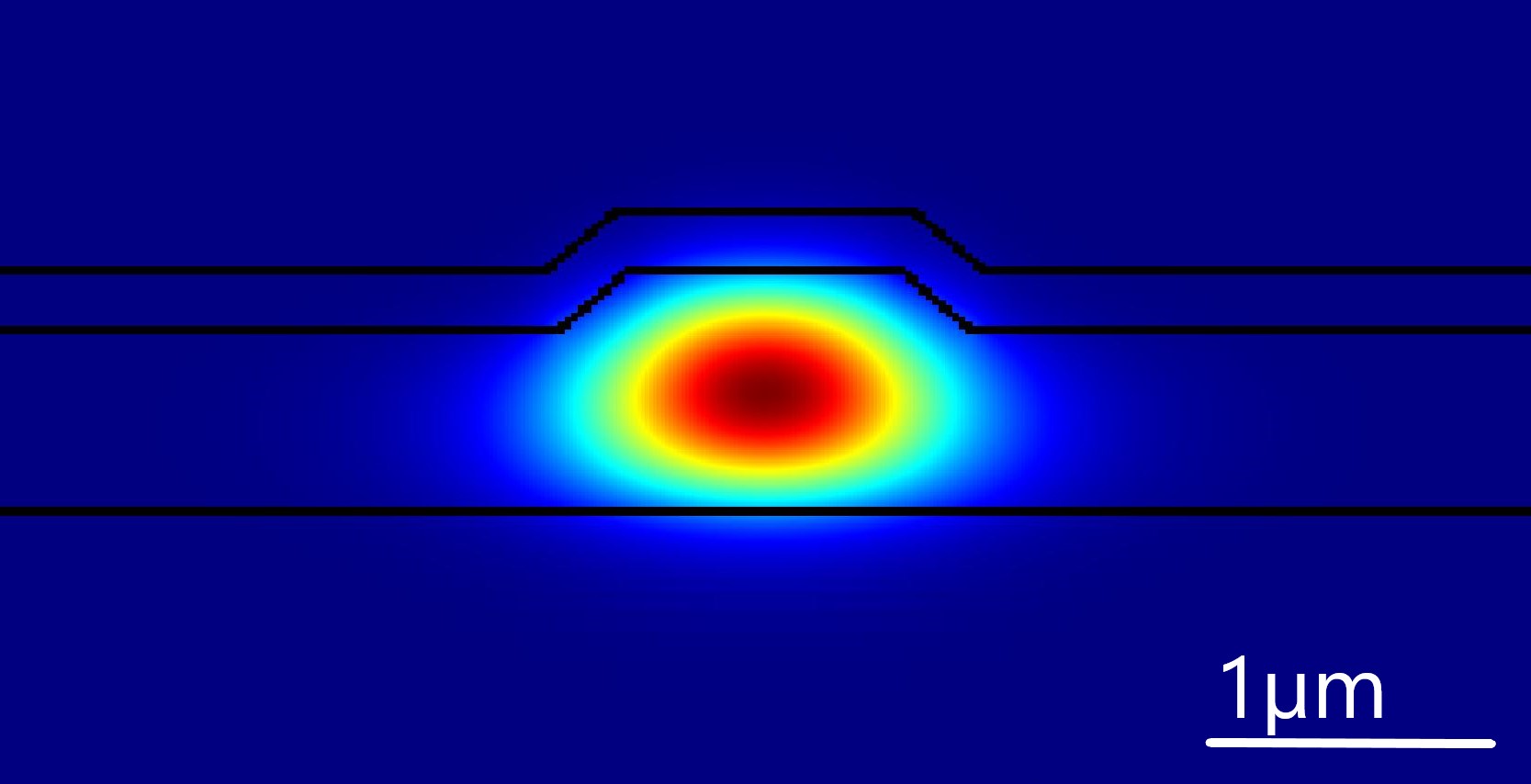}
         \caption{Single waveguide mode.}
         \label{Singlewaveguide}
     \end{subfigure}
     \hfill
     \begin{subfigure}[b]{0.3\textwidth}
         \centering
         \includegraphics[width=\textwidth]{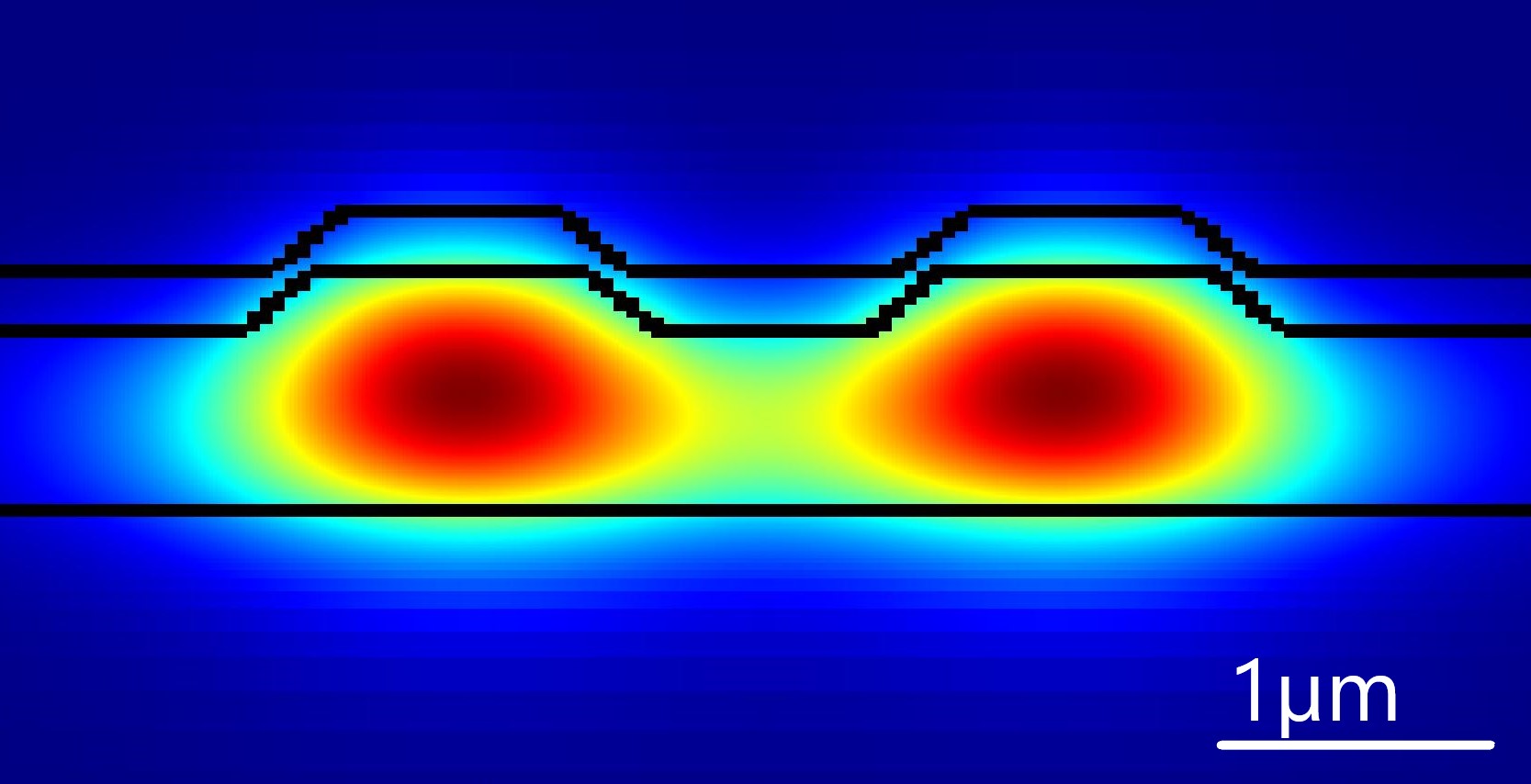}
         \caption{Symmetric mode.}
         \label{SymMode}
     \end{subfigure}
     \hfill
     \begin{subfigure}[b]{0.3\textwidth}
         \centering
         \includegraphics[width=\textwidth]{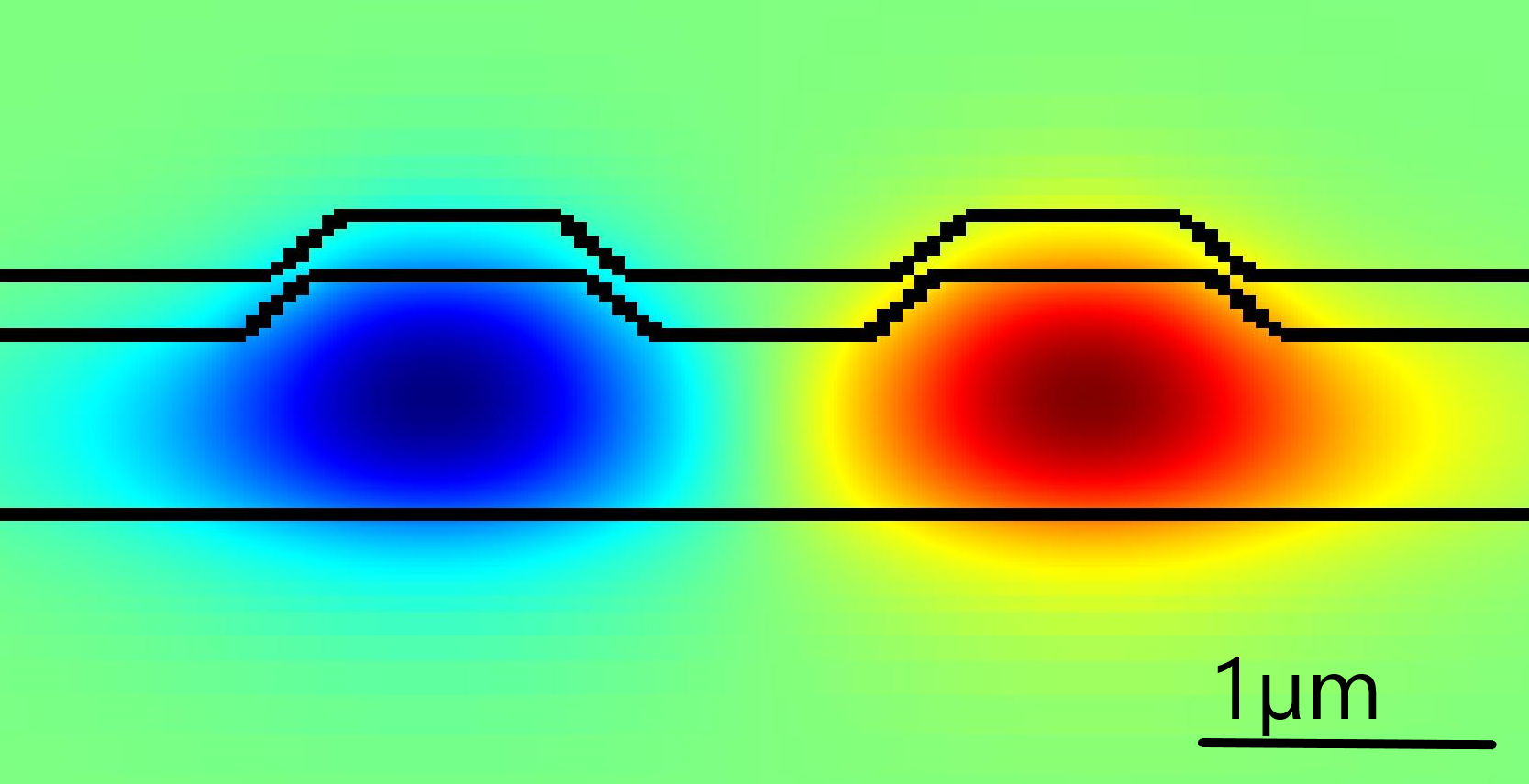}
         \caption{Anti-symmetric mode.}
         \label{AntisymMode}
     \end{subfigure}
        \caption{Simulations of waveguide structures in LNOI. The three images show the model of a single waveguide and of the directional coupler. The parameters of the waveguide that can be changed are the thin-film thickness, the etching depth, the top width, the angle and the thickness of the cladding layer and for the directional coupler the gap can be varied. For more information, see text.}
        \label{fig:three graphs}
\end{figure}
For that, we fixed the thin-film thickness to 600~nm and used X-cut LNOI. The sidewall angle of the waveguide is set to 60~° mirroring our fabrication process. Furthermore, the top width is adjusted to 1~\textmu m and we applied a cladding layer of SiO$_2$. For these parameters we simulated different etching depths to find a single mode waveguide geometry and chose 150~nm as an etching depth. In \autoref{Singlewaveguide} a single waveguide with the discussed geometry is shown. We can see the fundamental mode guided in the LN thin-film, the SiO$_2$ substrate underneath the LN and the SiO$_2$ cladding layer.\\ 
Afterwards, we simulated the coupling behaviour of a directional coupler and we introduced a second waveguide to our model defining the center to center distance between the two waveguides as a new parameter called gap $G$.
The calculated symmetric and anti-symmetric mode are shown in \autoref{SymMode} and \autoref{AntisymMode}. 
The resolution of the laser lithography used for defining the structures placed a lower limit on our gap width, which set a minimum coupling length for us. A gap width of 2.3 \textmu m yielded a coupling length of 137 \textmu m, leading to small device footprint well suited for high-density integration.
\begin{figure}[htbp]
\vspace*{-3mm}
\centering\includegraphics[width=7cm]{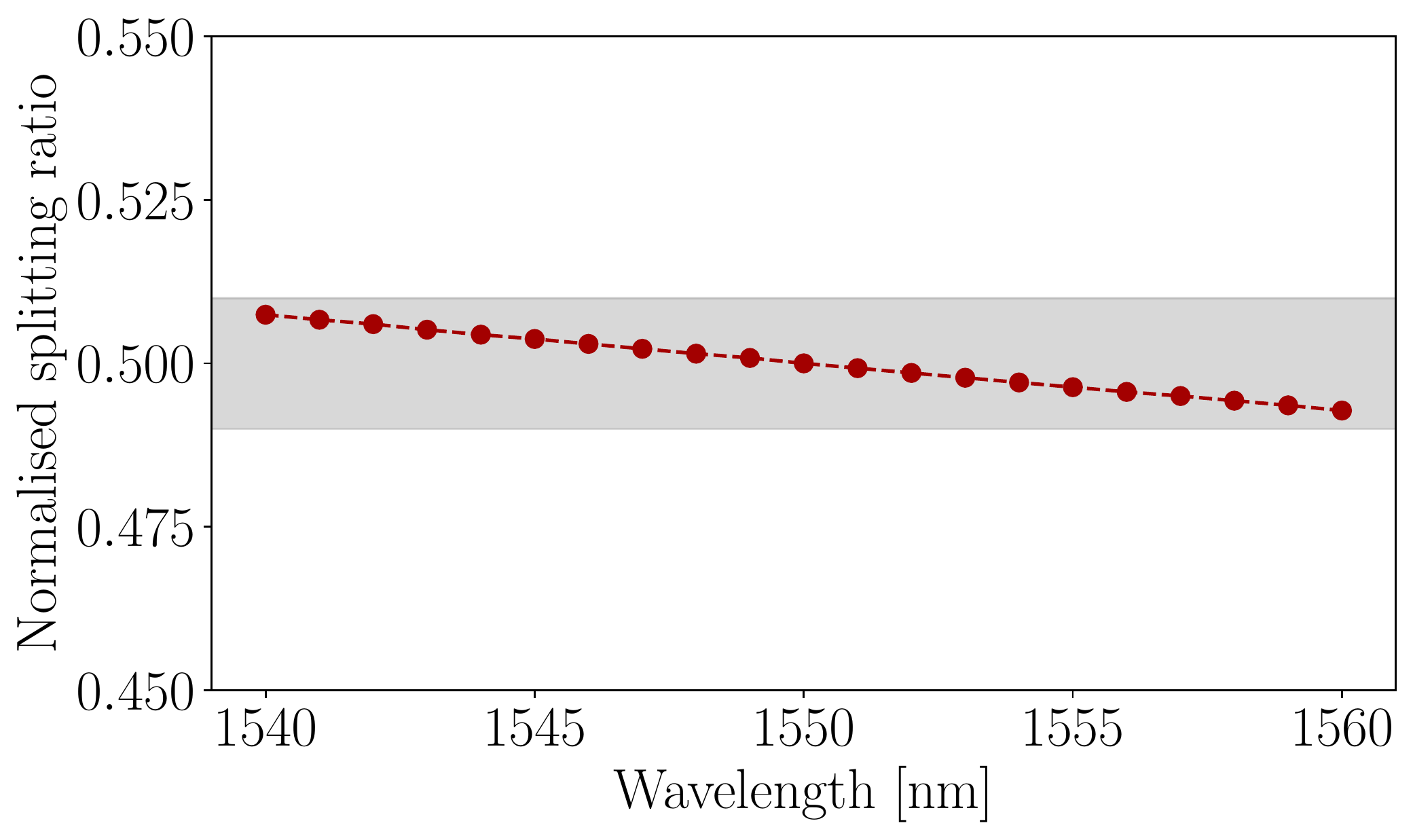}
\caption{The graph shows the normalised splitting ratio as a function of the wavelength. The splitting ratio lies within 1~$\%$ of the desired 50:50 value for wavelengths between 1540~nm and 1560~nm.}
\label{BandwidthSim}
\end{figure}
\\ For characterising the bandwidth of the coupler we show in \autoref{BandwidthSim} the normalised splitting ratio as a function of the wavelength. 
In a range of $\pm$ 10~nm the splitting ratio only varies by less than 1~$\%$ which shows that the coupler is wavelength independent for a broad range. 
Since the bandwidth of the single photons used for the HOM experiment is 1.8~nm at a wavelength of 1542.22~nm, we found a stable geometry with a large enough bandwidth for our experiment. 
In the next step, we fabricated this geometry to test the coupling behaviour.

\section{Fabrication of Directional Couplers} \label{Fabrication}
For the analysis of our directional couplers, we produced a sample with multiple waveguide groups consisting of a directional coupler and a straight waveguide to estimate the propagation losses. 
We changed the interaction length $L_I$ of the couplers from 30~\textmu m to 580~\textmu m to be able to determine the coupling length as explained in the next section.\\
We fabricate the waveguides via a physical dry etching process utilizing SiO$_2$ structured by a lift-off process as an etching mask. To realise the etching mask, we spin coat photo resist onto a LNOI sample (NanoLN), which is subsequently structured with a laser lithography system (Heidelberg Instruments DWL 66+). After developing, SiO$_2$ is deposited on the sample via sputtering (Prevac Sputtering system 518) and is structured by a lift-off process. Afterwards, the structured SiO$_2$ is used as an etching mask to transfer the structure in the lithium niobate thin-film layer. We use a dry etching process with a pure Argon plasma (Oxford Instruments PlasmalabSystem100). In a last step, the end-facet of the sample are chemo-mechanically polished. For this purpose, a  SiO$_2$ cladding layer is evaporated to protect the sample during polishing.

\section{Linear Characterisation of Waveguides and Directional Couplers} \label{Linear}

We measured the propagation losses of eight straight reference waveguides at a wavelength of 1550~nm using the so-called Fabry-Pérot method \cite{regener1985loss}. The light was coupled into and out of the LNOI waveguides with aspheric lenses. From these measurements, we estimate average propagation losses of ($4.85\pm0.95$)~dB/cm and note that these do not impede the quality of HOM interference, but do impact the overall measurement time.\\
Next, we investigated the coupling behaviour of the directional couplers. For that, we coupled laser light with a wavelength of 1550~nm (Santec TSL-550) into one input arm of the coupler and imaged the two output ports on a camera (Xenics Wildcat 640). 

\begin{figure}[b]
     \centering
     \begin{subfigure}[b]{0.45\textwidth}
         \centering
         \includegraphics[width=\textwidth]{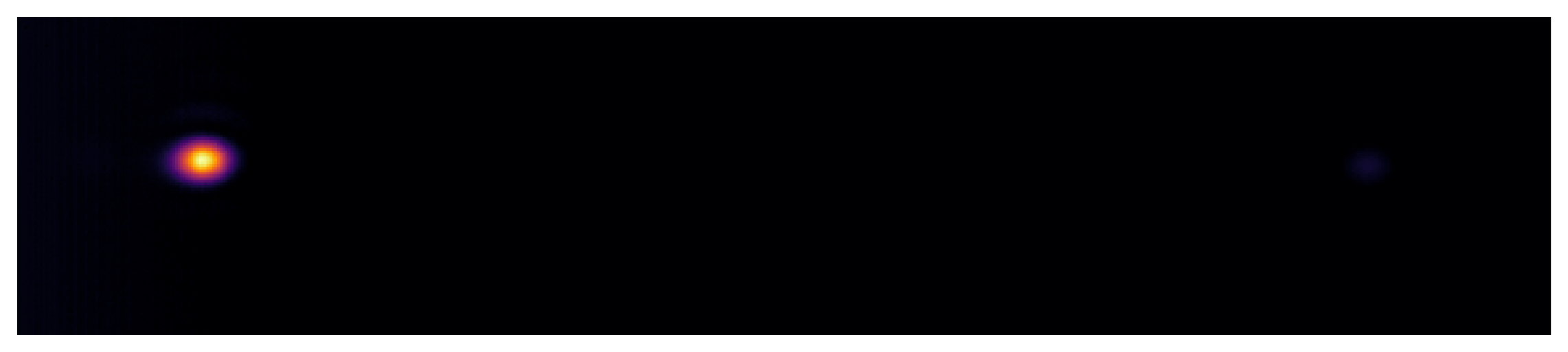}
         \caption{Interaction length: 58~\textmu m.}
         \label{coup58}
     \end{subfigure}
     \hfill
     \begin{subfigure}[b]{0.45\textwidth}
         \centering
         \includegraphics[width=\textwidth]{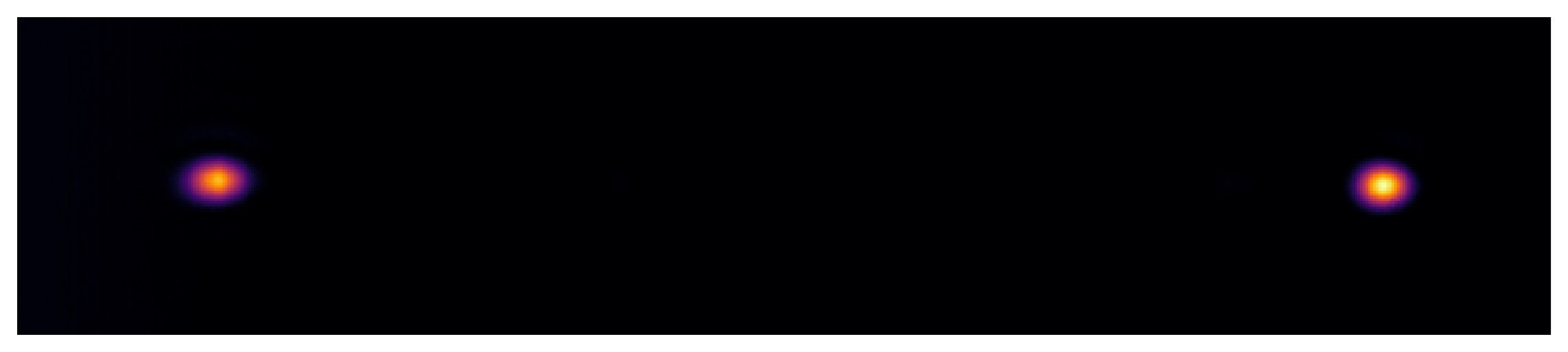}
         \caption{Interaction length: 257~\textmu m.}
         \label{coup257}
     \end{subfigure}
     \hfill
        \caption{Output modes of two directional coupler with different interaction length.\\ (a) All light coupled from one waveguide to the other. (b) A nearly 50:50 splitting ratio is reached.}
        \label{Modepictures}
\end{figure}

\noindent In \autoref{Modepictures} two examples of measured output modes can be seen. 
In \autoref{coup58} all the light from one input port coupled to the opposite output port. 
In contrast, in \autoref{coup257} the incoupled light is split nearly 50:50.
We then calculated for each output port the power ratio via this relation:

\begin{equation}
   P = \frac{P_1}{P_1 + P_2}.
\end{equation}

\begin{figure}[htbp]
\centering\includegraphics[width=7cm]{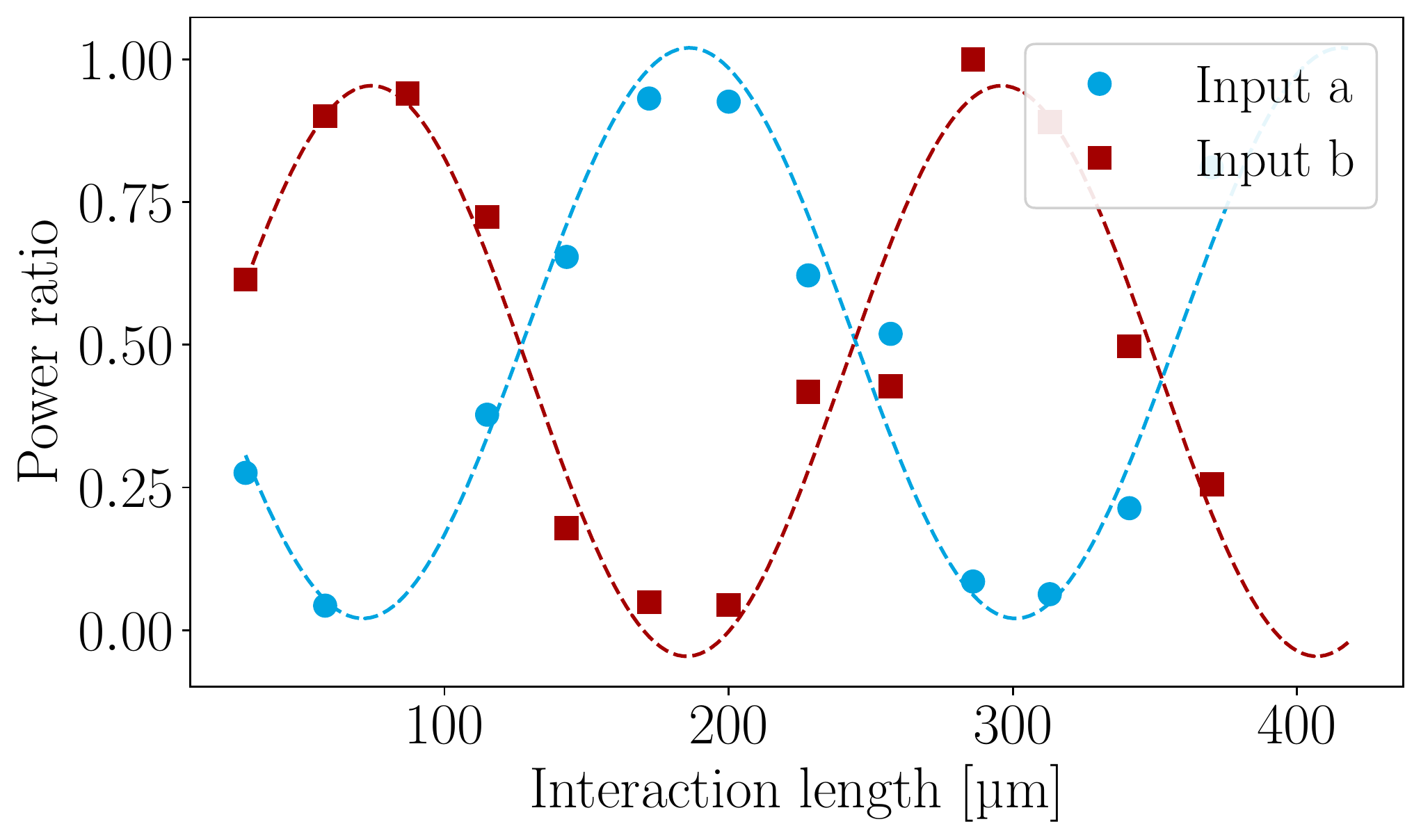}
\caption{Power ratio as a function of the interaction length. 
The measurement of the coupling behaviour shows that the light oscillates as expected between the two waveguides. 
The coupling length is (112.86$\pm$2.82)~\textmu m.}
\label{DCmeasuredcoupling}
\end{figure}

\noindent Here, $P_{1,2}$ is the power of the output mode 1 and 2, respectively.
In Figure \ref{DCmeasuredcoupling} the power ratio is plotted against the interaction length of the directional coupler for incoupling into input $a$ and input $b$. 
The offset for zero coupling length is due to coupling in the bendings leading to and from the coupling region. This effectively reduces the coupling length required for 50:50 splitting.\\
We extract the coupling length of the directional coupler from a sinusoidal fit, which yields lengths of ($114.85\pm0.04$)~\textmu m and ($110.87\pm0.04$)~\textmu m for inputs $a$ and $b$, respectively. The slight discrepancy can be caused by inhomogeneities in the fabrication. The average coupling length is ($112.86\pm2.82$) ~\textmu m, which is in excellent agreement with the expected 120 ~\textmu m predicted from the model. \\
For the HOM experiment a 50:50 coupler is needed. 
For this we choose the coupler that is nearest to this splitting ratio which is the one with the interaction length of 257~\textmu m. Note that this implied that we use a higher-order coupling process (light couples to and fro once), which is expected to have a reduced spectral bandwidth compared to a 0-th order coupling.

\section{Hong-Ou-Mandel Interference Experiment} \label{Quantum}

\begin{figure}[htbp]
\vspace*{-3mm}
\centering\includegraphics[width=\textwidth]{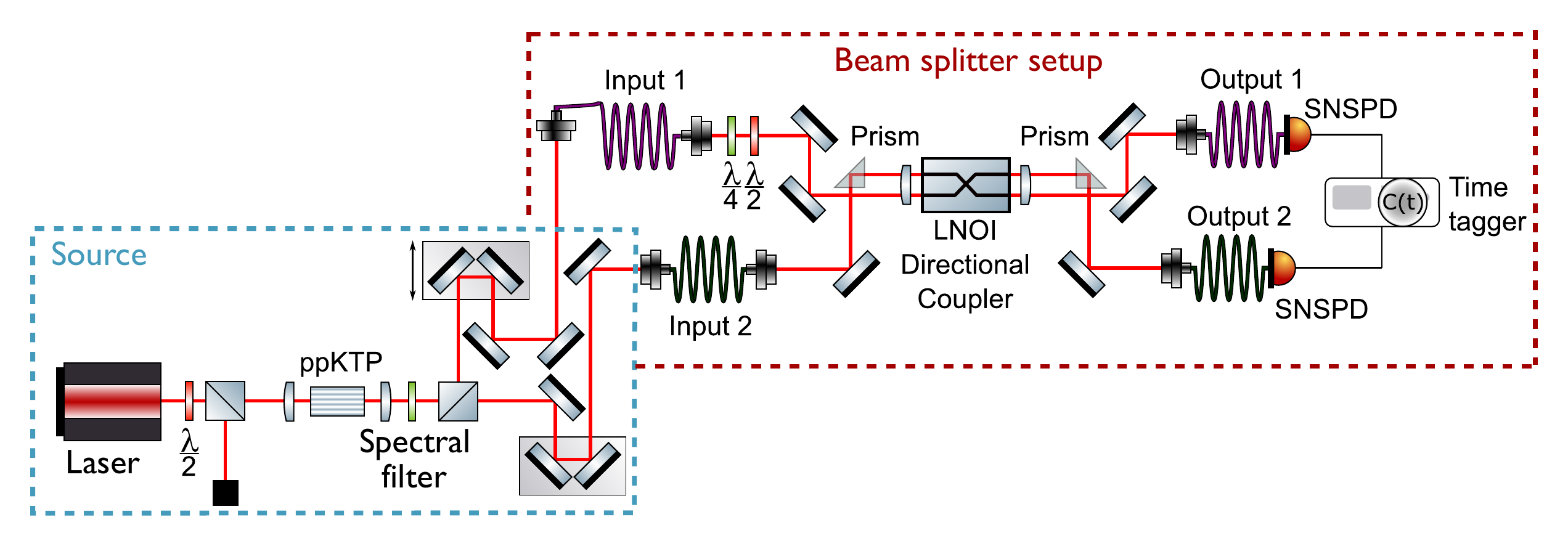}
\caption{Setup for the quantum experiment. The setup consists of two main parts. The first part is the preparation of the single photons and the second part consists of the LNOI BS.}
\label{Setup}
\end{figure}

\begin{figure}[htbp]
\centering\includegraphics[width=7cm]{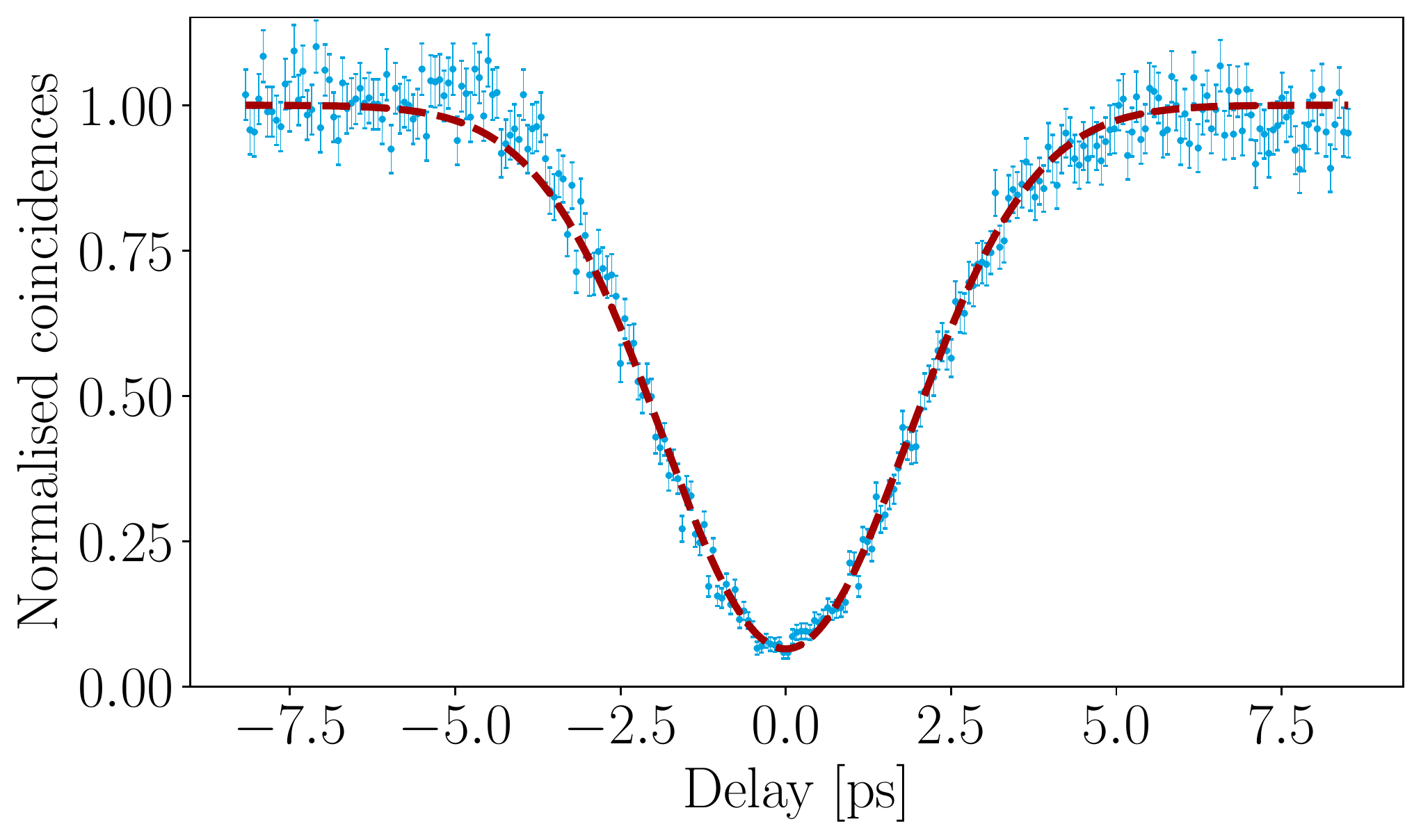}
\caption{Measured Hong-Ou-Mandel interference of the signal and idler photons interfered at the LNOI directional coupler. The Gaussian fit $1-0.935\cdot \text{exp}(-0.143 \cdot t^2)$ yields a visibility of ($93.5\pm0.7$)~$\%$.}
\label{Dip}
\end{figure}

For our final quantum experiment of the HOM interference, we realized a novel set-up as depicted in \autoref{Setup}.
The setup consists of two main parts. 
The first one is the source of the single photons generated by a spontaneous parameteric down-conversion (PDC) process in a ppKTP waveguide, and the second part is the LNOI BS.
The state preparation of the photons is the same as in \cite{nitsche2020local}. 
A type-II PDC process generates two photons with orthogonal polarisations at a wavelength of 1542.22~nm. The pulsed pump laser has a wavelength of 772.5~nm with a bandwidth of $\approx$0.3~nm.
We place a narrow band spectral filter with a FWHM of 1.8~nm after the ppKTP sample to suppress background radiation. 
We set the mean photon number of the generated PDC to a regime below 0.01 pairs per pulse to reduce the generation of higher photon components which would reduce the visibility. 
After the ppKTP sample the two photons are separated with a polarising BS and sent to delay lines. 
After the delay line each photon is coupled into a polarisation maintaining fiber and those fibers route to the BS setup where we couple the two photons from fibers to free space. 
The polarisation of one of the photons is controlled with a half-wave and quarter-wave plate to ensure indistinguishability.\\
The photons are coupled into the two inputs of the LNOI BS using a single aspheric lens. After the BS, we collect the photons with another lens and couple them to two standard SMF-28 single mode fibers. 
These fibers are connected to two superconducting nanowire single photon detectors (SNSPD) with a 70~ns dead time and efficiencies >95~$\%$ which are connected to a time tagger. 
To measure the HOM dip, we scan the delay line in 1\textmu m steps for 250~\textmu m. 
We perform several scans and add up the coincidences, to ensure that longterm intensity fluctuations are not affecting the HOM dip. We extract the visibility from a Gaussian fit (see \autoref{Dip}) and obtain a visibility of ($93.5\pm0.7$)~$\%$, where we did not subtract any background counts. Please note that the data shown in \autoref{Dip} has been normalised such that the baseline of the Gaussian fit is equal to 1 for better visualisation. 

\noindent Finally, we calculate the expected maximum visibility. 
From our model (see \autoref{DCmeasuredcoupling}), we extract an average splitting ratio of (54.6$\pm$3.8)~$\%$, slightly off from perfect balanced splitting. 
This limits the HOM visibility to (98.32$\pm$2.8)~$\%$. 
Further, we measured the HOM visibility of the source by replacing the LNOI BS with a tunable fibre BS and find a HOM visibility of (98.01$\pm$0.24)~$\%$. 
Combining the source visibility with the imperfect splitting ratio of the LNOI coupler, we obtain a maximum expected visibility of (96.36$\pm$2.8)~$\%$. Compared with our measured visibility of (93.5$\pm$0.7)~$\%$, we have a deviation of 3~$\%$. This could be due to effects that occur in high confined waveguide structures such as hybrid modes which have already been witnessed in LNOI \cite{kaushalram2020mode}. These hybrid modes can have an influence on the polarisation of the photons which would alter the splitting ratio of the devices and thus reduce the visibility of the dip. However, this effect is challenging to quantify. Note that even though we did not reach the maximal visibility, the measured visibility agrees with the expected one within the uncertainty range.
This demonstrates that the LNOI BS does not adversely effect HOM interference. 

\section{Conclusion} \label{Conclusion}

In this paper, we demonstrated HOM interference with telecom photons on LNOI, the missing piece for establishing LNOI as quantum-ready platform for integration. 
For this, we fabricated a directional coupler based on numerical simulations.
After that, we estimated the losses and the coupling behaviour. 
Finally, we performed a HOM experiment with single photons generated via a PDC process in ppKTP. 
With a measured visibility of (93.5$\pm$0.7)~$\%$ we can conclude that LNOI is a favourable platform for integrated quantum optics. 
Connecting several of these directional couplers to a sophisticated network and combining it with already shown single photon sources and electro-optical modulators in LNOI  opens the possibility for a quantum processor in LNOI.\\

\begin{backmatter}
\bmsection{Funding}
Deutsche Forschungsgemeinschaft (SFB-Geschäftszeichen TRR142/3-2022, Projektnummer 231447078.); Max Planck School of Photonics.

\bmsection{Acknowledgments}
Funded by the Deutsche Forschungsgemeinschaft (DFG, German Research Foundation) – SFB-Geschäftszeichen TRR142/3-2022 – Projektnummer 231447078. Laura Padberg and Silia Babel are part of the Max Planck School of Photonics supported by the German Federal Ministry of Education and Research (BMBF), the Max Planck Society, and the Fraunhofer Society.

\bmsection{Disclosures}
The authors declare no conflicts of interest.

\end{backmatter}


\bibliography{Optica-template}

\end{document}